\documentclass[preprint2]{aastex}
\newcommand{\ifpp}[1]{#1}
\newcommand{\ifms}[1]{}
\newcommand{\etal}{{\rm et~al.~}}

\begin{document}
\shortauthors{Bowyer \etal}
\shorttitle{EUV Emission from the Fornax Cluster}

\title{EXTREME ULTRAVIOLET EMISSION IN THE FORNAX CLUSTER OF 
GALAXIES}

\author{Stuart Bowyer, Eric Korpela and Thomas Bergh\"ofer}

\affil{Space Sciences Laboratory, University of California, Berkeley, CA
  94720-7450, USA}

\email{bowyer@ssl.berkeley.edu}

\begin{abstract}
We present studies of the Extreme Ultraviolet (EUV) emission in the Fornax 
cluster of galaxies; a relatively nearby well-studied cluster with X-ray emitting cluster 
gas and a very large radio source. We examine both the large-scale ($\sim$ size of the X-ray emitting cluster gas), and the small-scale ($<$ arc min) emission.  
We find that this cluster has large-scale diffuse EUV emission.  
However, at the sensitivity level of the existing EUVE data,
this emission is due entirely to the low energy tail of the X-ray emitting gas. 
We have also examined small-scale structures in raw EUVE images of this cluster.
We find that 
small-scale
irregularities are present in all raw Deep Survey images as a result of 
small-scale detector effects. These effects can be removed by appropriate 
flat-fielding.  
After 
flat-fielding, the Fornax cluster still shows a few significant regions of small-scale EUV
enhancement.  We find that these are emission from 
stars and galaxies in the field.  We find that at existing levels of sensitivity,
there is no excess EUV emission in the cluster on either large or small scales.
\end{abstract}


\section{Introduction}

The Fornax cluster (Abell S 373) is a relatively poor cluster at a distance 25 Mpc.  
It is well studied in the X-ray, radio, and optical bands.  It contains the radio galaxy 
Fornax A which is well known for its giant radio lobes that extend almost a degree 
across the sky.  The brightest optical galaxy in the group is NGC 1399; an E1 
Galaxy located near the center of the cluster.  The cluster has an associated X-ray 
emitting gas that has been studied by a number of investigators, most recently by 
Jones \etal 1997.  They find the cluster gas has a mean temperature of 1.3 keV and a 
heavy element abundance of 0.6 with respect to solar.  The cluster X-ray emission 
is $\sim$ 36 arc minutes in diameter roughly centered on NGC 1399.  

We studied the Fornax cluster in the hope that it might shed light on the underlying 
source mechanism of the EUV emission found in some clusters of galaxies.  We 
were particularly interested in this cluster since Bergh\"ofer \etal (2000a) showed 
that the jet in M87 may have activated the EUV emission in the Virgo cluster.  
Although Fornax A is well away from the cluster center and was not in the field 
covered by our observation, we entertained the possibility that this radio source 
might be activating processes in the central part of the cluster.  Throughout this 
paper we assume a Hubble constant of 50 km s$^{-1}$ Mpc$^{-1}$ and $q_o$= 0.5.

\section{Data and Analysis}

The Fornax cluster was observed from August 29th through September 2nd, 1998.   
During this period, 104 ks of data were obtained.  The cluster center was placed 
about 6 arcmin away from the known dead spot of the detector. 

\ifpp{
\begin{figure}[t]
\plotone{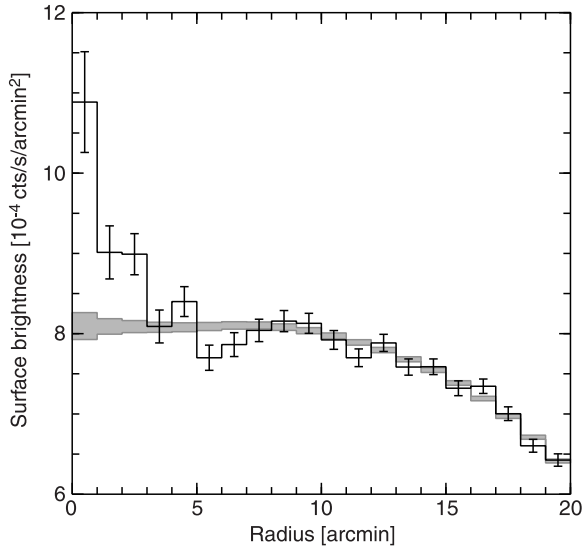}
\caption{\footnotesize The azimuthally averaged radial EUV emission profile of the Fornax cluster 
(solid line and one-$\sigma$ error bars). The azimuthally averaged radial profile
of the background, or flat-field, obtained from 788 ks of 
blank field data is shown as gray shaded regions. The one-$\sigma$ errors in this background are indicated by the size of the shaded area.}
\end{figure}
}

The reduction of the data was carried out with the EUVE package built in IRAF.  
We employed the analysis methods described in detail in Bowyer, Bergh\"ofer and 
Korpela 1999.  (See Bergh\"ofer, Bowyer, and Korpela 2000b for a definitive 
discussion of the validity of these procedures). Briefly, corrections for dead time 
and telemetry limitations were applied to the data set and a raw EUV image was 
produced. A flat non-photonic background determined from highly obscured 
regions at the outer most parts of the field was subtracted from this image.
We 
then computed the azimuthally averaged radial emission profile of the raw 
data centered on 
the cluster center. In Figure 1 we show this profile.
We 
also show the azimuthally averaged radial profile of the telescope sensitivity map (or flat-field) constructed from 788 ks of 
blank field observations. The flat-field profile and its statistical errors are shown as gray 
shaded regions. As can be seen from a comparison of these data sets, no EUV 
emission is detected at radii larger than 5 arcmin and the detection between 4 and 5 
arcmin is marginal at best. The statistical uncertainties in the flat-field are small 
because of the large number of counts in this data set.

We determined the EUV contribution of the low energy tail of the X-ray emitting 
cluster gas by analyzing 53,100 seconds of ROSAT PSPC archival data on this 
cluster. A reanalysis of this data using an accurate Galactic hydrogen column and 
appropriate interstellar absorption was necessary to obtain the correct EUV 
emission from the cluster, although the use of these improved parameters do not 
affect previous X-ray cluster gas measurements. We used the temperature of the X-
ray gas as obtained by Jones \etal 1997 and employed the MEKAL plasma code to 
derive the ROSAT PSPC to EUVE DS counts conversion factors for each separate 
radial bin. The Galactic hydrogen column employed was N (H) = $1.61 \times 10^{20} $cm$^{-2}$
(Murphy \etal 2000). The ISM absorption we employed is described in detail
in Bowyer \etal 1999.  This reference includes an extensive discussion of
the necessity of employing an improved EUV ISM absorption cross section in the
analysis of EUVE data.

\begin{figure}[t]
\plotone{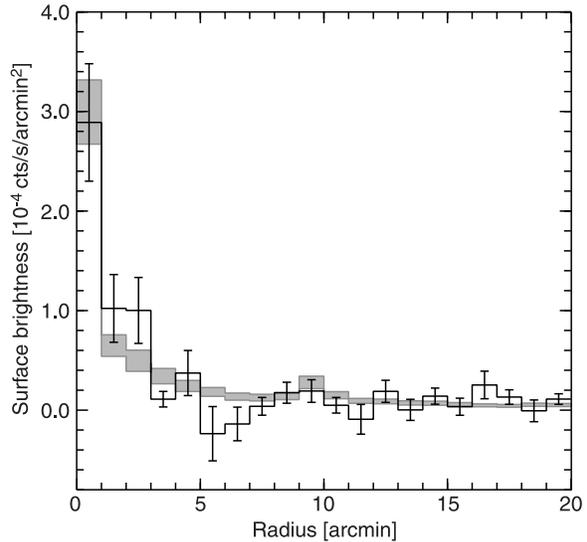}
\caption{\footnotesize The EUV emission in the Fornax cluster as derived from the data displayed 
in Fig. 1 (solid line with error bars). The statistical uncertainties in the flat-field and 
the signal are added in quadrature. We also show the EUV emission from the X-ray 
gas and its uncertainties as gray shaded regions. There is no evidence for excess 
large-scale EUV emission in the cluster.
}
\end{figure}

We established that the ROSAT PSPC to EUVE Deep Survey counts conversion 
factor fell between 220 (kT = 0.88 keV) and 125 (kT = 1.32 keV). We estimate the 
uncertainties in these values to be +/- 20\%.  Employing these values and using the 
azimuthally averaged radial X-ray emission profile derived from the PSPC hard 
energy band (0.5-2.4 keV), we derived upper and lower limits for the EUV 
emission from the X-ray emitting gas in the EUVE Deep Survey bandpass.  In 
Figure 2 we show the EUV emission from the X-ray emitting gas as shaded regions 
with uncertainties in this emission indicated by the size of the shaded bin.  We also 
show the EUV emission in the cluster as derived from the subtracting the 
flat-field shown in Figure 1 from the raw Fornax data shown in this figure.  
The errors in the signal and flat-field have been added in quadrature. The result shows that 
all the EUV flux in the cluster (at the sensitivity level of this observation) is produced by the X-ray cluster gas.

Using a 15.3 ks ROSAT observation of the Fornax cluster, Rangarajan et 
al. (1995) concluded that a
soft X-ray excess was needed to fit the data.  Their proposal of a 
$10^6$ K thermal component
with an HI column density of $1.61\times 10^{20}$ cm$^{-2}$
would produce an EUVE count rate of $1.8\times 10^{-3}$ 
s$^{-1}$. This
compares with the total emission we measure of   
$5.2\pm 1.3 \times 10^{-3}$ s$^{-1}$ and the expected count rate due
to the X-ray emitting gas of $4.25 \pm 0.25 \times 10^{-3}$ s$^{-1}$.
There is no evidence in the EUVE data that this excess emission is present,
although we can only exclude its existence at the 80\% confidence 
level.   Jones et al (1997) carried out an extensive analysis of the Fornax
cluster with a substantially longer ROSAT exposure of 54 ks (27.8 ks
usable).  They concluded that no soft excess was present, but a higher surface 
brightness was manifest in the innermost region. They suggested this was the 
result of a central cooling flow.  Similar effects have been found in other 
clusters with cooling flows, and this
is now the generally accepted explanation for this effect.

A detailed examination of the raw EUV image showed small-scale, 
$<$ arc minute, regions which deviate substantially from the mean. If these
features were intrinsic to the cluster this would provide important information
on the EUV cluster excess phenomenon.  
%In fact, Lieu, 
%in an oral presentation at the Ringburg Workshop on Diffuse Thermal and 
%Relativistic Plasma in Galaxy Clusters (Freyberg, April 1999), showed similar 
%small-scale features in the diffuse EUV emission in Abell 1795, and argued 
%that they were real features in the cluster that provided clues as to the 
%nature of the underlying source mechanism. 
Because of the importance of 
the possibility that such structures are present in one or more clusters, 
we have examined the small-scale structure we found in 
the Fornax image in detail. 

\begin{figure}[t]
\plotone{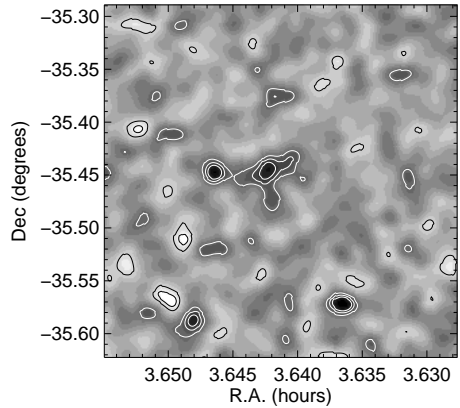}
\caption{\footnotesize A 20 arcmin by 20 arcmin map of the central portion of the 
Fornax cluster showing small-scale structure in the background subtracted
data.  
The contour levels are -3$\sigma$,
-2$\sigma$, +2$\sigma$, +3$\sigma$, and +4$\sigma$.
The largest positive deviation is coincident with the X-ray cluster gas.
The point 
source to the southeast is the galaxy NGC 1404.  The position of the source
to the southwest is coincident with an unnamed star.
The position of the point source to the east of the cluster center is 
consistent with the galaxy CGF 1-3. }
\end{figure}

The pixel scale of our Fornax data set was 4.6 arc seconds. We smoothed 
this data in 
order to increase the number of counts in individually resolved features 
using a 1 
arc minute FWHM Gaussian; this is about the 90\% included energy width of the EUVE 
DS Telescope. We then subtracted our flat-field, similarly smoothed, from this 
image. The resultant image is shown in Fig. 3. The data in this image shows obvious 
deviations, both greater and less than the mean.  

We carried out a variety of tests to determine if these deviations were statistically 
significant, and to determine whether or not they were unique to the Fornax cluster. 
To this end we examined both our 104 ks of Fornax cluster data, and a number of 
blank field data sets. Hardcastle 2000 (and references therein) has shown that 
extreme care must be used in determining the validity of features in an image that 
has been smoothed as described. The statistics in the smoothed image are neither 
Poissonian (because of the smoothing), nor Gaussian (because of the small number 
of counts in each cell).  Hardcastle points out that one cannot simply determine the 
r.m.s. dispersion of an image after smoothing, multiply by n, add the mean, and call 
the resulting contour level ``$n\sigma$".  
%Using his terminology, a ``$3\sigma$" contour 
%includes all but 0.135% of the bins in a smoothed random (Poissonian) data set.  
%This is in contrast to equating ``$3\sigma$" to ``3 standard deviations"; the fraction of 
%a smoothed random data set exceeding ``3 standard deviations" is very likely to be 
%larger than 0.135%.  

Hardcastle points out that there is no analytic 
solution to the problem of  establishing 
the statistical significance of a data set after smoothing. Hardcastle 
describes a Monte Carlo procedure that gives valid results. A field of simulated 
Poisson noise with the same bin size as the true data set is convolved with a 
Gaussian used with this data, and the statistical uncertainty levels are derived 
directly from the distribution of the resulting noise. 
An ``equivalent" $3\sigma$ contour is defined as the contour that includes all but 
0.135\% of the data.  Contours of other significance levels are defined in an
analogous manner.

Following these procedures, we 
carried out a Monte Carlo simulation of Poisson noise in an empty field to determine the expected 
fraction of the field exceeding various levels of significance. The results of this simulation are 
shown in Column 1 of Table 1.  Given that this data set is used to define
these levels, they precisely match their respective statistical level.
For comparison we also show  in Column 2 the expected 
deviations (at these equivalent sigma levels) as calculated from Poisson statistics. These are clearly 
different, confirming the work of Hardcastle. In Column 3 we show the fraction of 
regions in the Fornax field with these statistical significance levels.
The Fornax image shows more positive deviations
than would be present by chance.

To further evaluate these fluctuations, we 
examined a number of blank field data sets. In Column 1 of Table 2 we show the 
fraction of the field exceeding various levels of significance in a set of 205 ks of 
blank field data (set 1). In this column we list the results for the blank field data after 
subtraction of the mean. There are substantial regions  showing statistically
significant structure in this raw data set.  In Column 2 we show the results after subtraction of an 
independent set of 425 ks of blank field data (set 2) scaled in the same manner 
used with the cluster data. This subtraction provides a flat-fielded image of the set
1 blank field data.  In Column 3 we show the Monte Carlo simulation of smoothed
Poisson
noise (reproduced here from Column 1 of Table 1 for the convenience of the reader.)
The fluctuations in the flat-fielded blank field  data are consistent with the Monte Carlo
simulation of smoothed Poisson noise.

\section{Discussion and Conclusions}

The data displayed in Figure 1 shows the Fornax cluster exhibits large-scale diffuse 
EUV emission.  However, the results displayed in Figure 2 show this emission is 
due entirely to the low energy tail of the X-ray emitting gas and that there
is no detectable excess EUV flux in the central region of the cluster. 
The 
Fornax A radio 
source is not activating EUV emission in the central part of the cluster
at the sensitivity level provided by the existing data. 

The question as to what activates the production of large-scale EUV emission in 
some cluster of galaxies remains unanswered. A major factor limiting this
inquiry is that few of the 
sources that have been studied in an appropriate manor have been found to exhibit 
this emission, hence few clues are available as to the nature of the underlying source 
mechanism(s). Excess EUV emission is clearly present in the Coma cluster 
(Bowyer \etal, 
1999,
Korpela \etal, in progress).   
It is also present in the Virgo cluster 
(Bergh\"ofer \etal 2000a).  Intriguingly, the character of the emission in
the Virgo cluster emission
is quite different than that in the Coma cluster. The only generalization that 
can be made from a study of the emission in these two clusters is that the emission is 
not the product of a gravitationally bound gas. Excess EUV emission has been
claimed to have been detected in Abell 1795 (Mittaz \etal 1998) and 
Abell 2199 (Lieu \etal 1999).   However, these authors used a theoretically
derived flat flat-field in the analysis of these clusters which is now 
acknowledged by all researchers to be inappropriate.
An appropriate analysis of
Abell 1795 and Abell 2199 (Bowyer \etal 1999)
does not 
show excess EUV emission.  

We next discuss the small-scale structure observed
in the raw Fornax image. Our analysis of blank field data
shows that any field in which a constant level has been
subtracted from the raw data will show a substantial number of small-scale
features.  These will be positive, or negative, or a mix, depending upon the
numerical
value of the constant
level which has been subtracted.  However, a comparison of the flat-fielded blank field
data in Column 2 of Table 2 with the true significance levels shown 
in Column 3 show 
the fluctuations in a correctly flat-fielded blank field image are 
consistent with random noise.

We turn now to the question of small-scale structure in the correctly 
flat-fielded Fornax image. Most of the small-scale
features seen in the raw data are no longer present, consistent with our
demonstration that most of this structure is due to detector effects.  However,
a few regions with
statistically significantly positive deviations are present even after 
appropriate 
flat-fielding has been carried out. The large-scale feature in the raw EUVE
data is
closely aligned with the X-ray emitting cluster gas, and is clearly due to
the EUV emission of this gas. It can be eliminated by subtracting a properly
scaled X-ray image.  
The next most significant features have widths $\simeq$ the point spread 
function of the Deep Survey Telescope; they are associated with known 
galaxies and an unnamed star, and have nothing to do with intrinsic cluster 
emission. It is highly likely that the few remaining enhancements in this 
field are also due to galaxies in the Fornax cluster or to unidentified 
field stars. The negative fluctuations are consistent with their being 
the result of chance alignments of random fluctuations in either the 
Fornax data set or the background data set. We conclude there is no 
evidence for small-scale EUV enhancements or deficits in the Fornax cluster. 

Finally, we emphasize that investigators of diffuse emission with EUVE (or 
other spacecraft) should be aware of the complexities in evaluating 
apparent small-scale fluctuations in the raw data set.

\section{Acknowledgments}

 This work was supported in part by NASA cooperative agreement NCC5-138 and 
an EUVE Guest Observer Mini Grant.  TWB was supported in part by a Feodor-
Lymen Fellowship of the Alexander-von-Humboldt-Stiftung.  We thank the 
anonymous referee for useful comments which greatly improved this paper.

\appendix{}
\begin{table}[t]
\begin{center}
\begin{tabular}{lrrr}
\hline
  \multicolumn{4}{c}{Fraction of Field Exceeding Significance Levels} \\
  & \multicolumn{1}{c}{Monte Carlo Simulation} & \multicolumn{1}{c}{Poisson Statistics} & \multicolumn{1}{c}{Fornax Field} \\
\hline\hline
$>4\sigma$ &	0.00003 &	0.003 &	0.003 \\
$>3\sigma$ &    0.0013 &   0.010  &       0.007 \\
$>2\sigma$ &       0.023   &   	0.066 &	0.031 \\
$>1\sigma$ &       0.159 &	0.228 &        0.161 \\
$<-1\sigma$ &       0.159  &        0.217 &    	0.173 \\
$<-2\sigma$ &       0.023 & 	0.062 &	0.024 \\
$<-3\sigma$ &	0.0013 &	0.010 & 0.0019 \\
\hline
\end{tabular}
\caption{Statistical Validity of the Small Scale ($<$ $\sim$ arc min) Structure in the
Fornax
Cluster
}
\end{center}
\end{table}

\begin{table}[b] 
\begin{center}
\begin{tabular}{lrrr}
\hline
  \multicolumn{4}{c}{Fraction of Field Exceeding Significance Levels} \\
  & \multicolumn{1}{c}{Blank Field 1-Mean} & \multicolumn{1}{c}{Blank Field 1-Blank Field 2} & \multicolumn{1}{c}{Monte Carlo Simulation} \\
\hline\hline
$>4\sigma$ & 0.0000 & 0.0000 &	0.00003 \\
$>3\sigma$ & 0.005  & 0.0010 &  0.0013 \\
$>2\sigma$ & 0.074  & 0.025 &     0.023  \\
$>1\sigma$ & 0.267  & 0.159 &      0.159 \\
$<-1\sigma$ & 0.245 & 0.162 &      0.159 \\
$<-2\sigma$ & 0.094 & 0.024 &      0.023 \\
$<-3\sigma$ & 0.025 & 0.0018 & 	0.0013 \\
$<-4\sigma$ & 0.010 & 0.0007 &	0.00003 \\
\hline
\end{tabular}
\caption{
Statistical Validity of the Small Scale ($<$ $\sim$ arc min) Structure in Blank Field 
Data
}
\end{center}
\end{table}

\ifms{
\newpage
\figcaption[bowyer_fig1.ps]{The azimuthally averaged radial EUV emission profile of the Fornax cluster 
(solid line and one-$\sigma$ error bars). The azimuthally averaged radial profile
of the background, or flat-field, obtained from 788 ks of 
blank field data is shown as gray shaded regions. The one-$\sigma$ errors in this background are indicated by the size of the shaded area.}

\figcaption[bowyer_fig2.ps]{The EUV emission in the Fornax cluster as derived from the data displayed 
in Fig. 1 (solid line with error bars). The statistical uncertainties in the flat-field and 
the signal are added in quadrature. We also show the EUV emission from the X-ray 
gas and its uncertainties as gray shaded regions. There is no evidence for excess 
large-scale EUV emission in the cluster.
}

\figcaption[bowyer_fig3.ps]{An 20 arcmin by 20 arcmin map of the central portion of the 
Fornax cluster showing small-scale structure in the background subtracted
data.  
The contour levels are -3$\sigma$,
-2$\sigma$, +2$\sigma$, +3$\sigma$, and +4$\sigma$.
The largest positive deviation is coincident with the X-ray cluster gas.
The point 
source to the southeast is the galaxy NGC 1404.  The position of the source
to the southwest is coincident with an unnamed star.
The position of the point source to the east of the cluster center is 
consistent with the galaxy CGF 1-3. }
}


\begin{thebibliography}{XXXXXX}
\bibitem[Bergh\"ofer \etal 2000a]{bbk2000a}
Bergh\"ofer, T.W., Bowyer, S., \& Korpela, E.J. 2000a, ApJ, 535,615

\bibitem[Bergh\"ofer \etal 2000b]{bbk2000b}
Bergh\"ofer, T.W., Bowyer, S., \& Korpela, E.J. 2000b, ApJ, in press

\bibitem[Bowyer \etal 1999]{bbk1999}
Bowyer, S., Bergh\"ofer, T.W., \& Korpela, E.J. 1999, ApJ, 526,592-598

\bibitem[Ekers \etal 1989]{ekers89}
Ekers, R.~D., Wall, J.W., Shaver, P.A., Goss, W.M.,
Fosbury, R.A.E., Danziger,I.J., Moorwood, A.F.M., Malin, D.F.,
Monk, A.S, \& Ekers, J.A. 1989, MNRAS, 236, 737-777

\bibitem[Geldzahler \& Fomalont 1984]{GF89}
Geldzahler, B. J., \& Fomalont, E. B. 1984, AJ, 89, 1650

\bibitem[Hardcastle 2000]{hard2000}
Hardcastle, M.J. 2000, A\&A, 357, 884.

\bibitem[Korpela \etal 2000]{korpela00}
Korpela, E.J., Bowyer, S., \& Bergh\"ofer T.W. 2000, in prep.

\bibitem[Lieu \etal 1999]{lieu99}
Lieu, R., Bonamente, M., \& Mittaz, J. 1999 ApJ, 517, L91

\bibitem[Jones \etal 1997]{Jones97}
Jones, C., Stern, C., Forman, W., Breen, J., David, L., Tucker, W., \& 
Franx, M. 1997, 
ApJ, 482, 143-155

\bibitem[Mills 1954]{mills54}
Mills, B.Y. 1954, The Observatory, 74, 248

\bibitem[Mittaz \etal 1998]{mittaz98}
Mittaz, J., Lieu, R., \& Lockman, F. 1998, ApJ 498, L17

\bibitem[Murphy \etal 2000]{murphy2000}
Murphy, E., Sebach, K., \& Lockman, F. 2000, ApJS in press

\bibitem[Rangarajan \etal 1995]{ranga95}
Rangarajan, F.V.N., Fabian, A.C., Forman, W.R., \& Jones, C. 1995 MNRAS, 272, 665.

\bibitem[Robertson \& Roach 1990]{rr90}
Robertson, J. \& Roach, G. 1990, MNRAS, 247, 387-399

\bibitem[Wade 1961]{wade61}
Wade, C.M. 1961, Pub. N.R.A.O., 1, 99

\end{thebibliography}
\end{document}